\documentclass[twocolumn,showpacs,aps,prd,]{revtex4-1}
\usepackage{graphicx}
\usepackage{dcolumn}
\usepackage{amsmath}
\usepackage{epsfig}
\usepackage{subfigure}
\usepackage{epstopdf}
\usepackage{multirow}
\usepackage{mathrsfs}
\usepackage[colorlinks,urlcolor=blue,citecolor=blue,linkcolor=blue] {hyperref}
\usepackage{lineno}
\usepackage{enumerate}
\usepackage{float}

\begin{document}
%\linenumbers
\title{\boldmath Study of the electromagnetic Dalitz decays $\psi(\Upsilon) \to \eta_{c}(\eta_{b}) l^{+} l^{-}$}

%\author{Li-Min Gu$^{1}$, Hai-Bo Li$^{2,3}$, Xin-Xin Ma$^{2,3,*}$ and Mao-Zhi Yang$^{4}$ \\
%$^{1}$Nanjing University, Nanjing 210093, People's Republic of China \\
%$^{2}$Institute of High Energy Physics, Beijing 100049, People's Republic of China \\
%$^{3}$University of Chinese Academy of Sciences, Beijing 100049, People's Republic of China \\
%$^{4}$ School of Physics, Nankai University, Tianjin 300071, China\\
%$^*$maxx@ihep.ac.cn\\
%}
\author{Li-Min Gu$^{1}$} \email{gulm@ihep.ac.cn}
\author{Hai-Bo Li$^{2,3}$} \email{lihb@ihep.ac.cn}
\author{Xin-Xin Ma$^{2,3}$} \email{maxx@ihep.ac.cn} 
\author{Mao-Zhi Yang$^{4}$} \email{yangmz@nankai.edu.cn}
\affiliation{$^{1}$Nanjing University, Nanjing 210093, People's Republic of China \\
$^{2}$Institute of High Energy Physics, Beijing 100049, People's Republic of China \\
$^{3}$University of Chinese Academy of Sciences, Beijing 100049, People's Republic of China \\
$^{4}$ School of Physics, Nankai University, Tianjin 300071, China
}

\date{\today}
\begin{abstract}
    We study the electromagnetic Dalitz decays, $\psi \to \eta_{c} l^{+}
    l^{-}$ and $\Upsilon \to \eta_{b} l^{+} l^{-}$ ($l = e$ or $\mu$), in which
    the lepton pair comes from the virtual photon emitted by the M1 transition
    from $c\bar{c}$ ($b\bar{b}$) spin triplet state to the spin singlet state.
    We estimate the partial width of $\psi(\Upsilon) \to \eta_{c}(\eta_{b})
    l^{+} l^{-}$, based on the simple pole approximation. Besides, based on
    different QCD models,  the partial width of $\psi(\Upsilon) \to
    \eta_{c}(\eta_{b}) \gamma$ is determined.
\end{abstract}

\pacs{13.20.Gd, 13.40.Hq, 12.38.Aw}
\maketitle

\section{\boldmath Introduction}
The electromagnetic (EM) Dalitz decays, $V \to P l^{+} l^{-}$, where $V$ and $P$
are vectors and pseudoscalar mesons, and $l$ denotes  lepton ($e$, $\mu$),
provide an ideal opportunity to probe the structure of hadronic states and to
investigate the fundamental mechanisms of the interactions between photons and
hadrons~\cite{Landsberg:1986sk, Landsberg:1986fd}. The lepton pair $l^{+}l^{-}$
comes from an off-shell photon,  radiated from the transition between $V$ and
$P$. Assuming pointlike particles, the process can be exactly described by
QED~\cite{Kroll:1955zu}. Otherwise, the structure-dependent partial width can be
modified by transition form factor $f_{\rm VP}(q^{2})$, which can be estimated
based on QCD models \cite{Achasov:1992ku, Klingl:1996by, Faessler:1999de,
Terschluesen:2010ik, Ivashyn:2011hb} and provides information of the EM
structure arising from the $V$-$P$ transition. The M1 transition between $\psi
(\Upsilon)$ and $\eta_{c} (\eta_{b})$ is widely studied in theory
\cite{Shifman:1979nx,Khodjamirian:1983gd,Beilin:1985da,Zhang:1991et,Ebert:2002pp,Lahde:2002wj,Hwang:2006cua,Dudek:2006ej,Ke:2010pp,Donald:2012ga,Becirevic:2012dc,Pineda:2013lta},
 and the M1 transition between $\psi$ and $\eta_{\rm c}$ has been observed with
 the average branching fraction, ${\cal B(}J/\psi \to \gamma\eta_{\rm c}(1S)) =
 (1.7 \pm 0.4) \%$~\cite{Tanabashi:2018oca}.
 In the following ratio of branching fractions:
\begin{equation}
    R \equiv \frac{B(\psi \to e^+ e^-  \eta_c)}{ B(\psi \to \gamma \eta_{c})},
\end{equation}
many theoretical uncertainties can cancel; therefore, it can be used to test theoretical models.
Experimentally, the EM Dalitz decays of light unflavored vector mesons
($\rho^{0}, \omega, \phi$) have been widely observed~\cite{Tanabashi:2018oca}
and several decays of charmonium vector mesons ($J/\psi, \psi^{\prime}$) to
light pseudoscalar mesons, which are studied in Ref.~\cite{Fu:2011yy}, have
been observed recently by BESIII experiment~\cite{Tanabashi:2018oca,
Ablikim:2018xxs,Ablikim:2014nro}. In Table~\ref{tab:summaryresult}, we summarize
the experimental results of the EM decays for the light unflavored vector mesons
($\rho^{0}, \omega, \phi$) and charmonium vector mesons ($J/\psi,
\psi^{\prime}$). The results indicate that the ratios of the EM Dalitz decay to
the corresponding radiative decays are suppressed by 2 orders of magnitude.

However, in previous paper~\cite{Fu:2011yy}, which focus on the EM Dalitz decays
of $J/\psi$,  it is assumed that the $\psi(\Upsilon)$ is totally unpolarized.  In
this paper, we investigate the polarization of $\psi(\Upsilon)$ produced in $e^+
e^-$ collisions, then  deduce the general form of the decay width, as a function
of polarization vector of  $\psi(\Upsilon)$, at last apply it to the
$\psi(\Upsilon) \to l^{+}l^{-} \eta_{c}(\eta_{b})$ decay to predict its
branching fraction.

\begin{table}[!htbp]
    \centering
    \caption{
        The branching fractions of EM Dalitz decays $V \to P l^{+} l^{-}$ and
        ratios of the EM Dalitz decays to the corresponding radiative decays of
        the vector mesons. These data are from PDG2018~\cite{Tanabashi:2018oca}.
    }
    \begin{tabular}{p{2.2cm}p{3.1cm}<{\centering}p{3.1cm}<{\centering}}
        \hline \hline
        Decay mode  &  Branching fraction  &  $\frac{\Gamma(V \to P l^{+} l^{-})}{\Gamma(V \to P \gamma)}$  \\
        \hline
        $\rho^{0} \to \pi^{0} e^{+}  e^{-}$  &  $<1.2 \times 10^{-5}$ &  $<2.6 \times 10^{-2}$   \\
        $\omega \to \pi^{0} e^{+}  e^{-}$  &  $(7.7\pm0.6) \times 10^{-4}$  &  $(0.91\pm0.08) \times 10^{-2}$    \\
        $\omega \to \pi^{0} \mu^{+}  \mu^{-}$  &  $(1.34\pm0.18) \times 10^{-4}$  &  $(0.16\pm0.02) \times 10^{-2}$     \\
        $\phi \to \pi^{0} e^{+}  e^{-}$  &  $(1.33^{+0.07}_{-0.10}) \times 10^{-5}$  &  $(1.02^{+0.07}_{-0.09}) \times 10^{-2}$    \\
        $\phi \to \eta e^{+}  e^{-}$  &  $(1.08\pm0.04) \times 10^{-4}$  &  $(0.83\pm0.03) \times 10^{-2}$    \\
        $\phi \to \eta \mu^{+}  \mu^{-}$  &  $<9.4 \times 10^{-6}$  &  $<0.07 \times 10^{-2}$    \\
        $J/\psi \to \pi^{0} e^{+} e^{-}$  &  $(7.6\pm1.4) \times 10^{-7}$   & $(2.18^{+0.45}_{-0.44}) \times 10^{-2}$  \\
        $J/\psi \to \eta e^{+} e^{-}$  &  $(1.16\pm0.09) \times 10^{-5}$  &  $(1.05\pm0.09) \times 10^{-2}$  \\
        $J/\psi \to \eta^{\prime} e^{+} e^{-}$  &  $(5.81\pm0.35) \times 10^{-5}$  &  $(1.13\pm0.08) \times 10^{-2}$   \\
        $\psi^{\prime} \to \eta^{\prime} e^{+} e^{-}$  &  $(1.90\pm0.27) \times 10^{-6}$   &  $(1.53\pm0.22) \times 10^{-2}$   \\
        \hline \hline
    \end{tabular}
    \label{tab:summaryresult}
\end{table}

\section{\boldmath General formula for the EM Dalitz decay $V \to P l^{+} l^{-}$}
The amplitude of the EM Dalitz decay, $V \to P l^{+} l^{-}$,  has been obtained
in the literature previously, which can be written in a Lorentz-invariant
form~\cite{Landsberg:1986sk,Landsberg:1986fd,Fu:2011yy}
\begin{equation}
T(V \to P l^{+} l^{-}) = 4 \pi \alpha f_{V\!P}\epsilon^{\mu\nu\rho\sigma}p_{\mu}q_{\nu}\epsilon_{\rho}\frac{1}{q^{2}}\bar{u}_{1} \gamma_{\sigma} \nu_{2},
\end{equation}
where $\alpha$ is the fine-structure constant, $f_{V\!P}$ the transition form
factor, $\epsilon^{\mu\nu\rho\sigma}$  the Levi-Civita tensor, $p_{\mu}$ the
momentum of  the pseudoscalar meson, and $q_{\nu} = k_{1}+k_{2}$ with $k_{1}$
and $k_{2}$ the momenta of the $l^{+}$ and $l^{-}$. After summing over the spin
of leptons, the amplitude squared is
\begin{equation}
    |T(V \to P l^{+} l^{-})|^{2} = 16\pi^{2}\alpha^{2}\frac{|f_{V\!P}(q^{2})|^{2}}{q^{4}} \cdot h ,
\end{equation}
where
\begin{equation}
\begin{split}
h = ~ &  8 m^{2}_{V} m^{2}_{l} (q^{2} \epsilon \cdot \epsilon^{*} - q \cdot \epsilon  q \cdot \epsilon^{*} ) \\
       & -2 m^{2}_{V} q^{4}(k_{1} - k_{2} )\cdot \epsilon (k_{1}  - k_{2}) \cdot \epsilon^{*}
      \\ & +8m^{2}_{l} q\cdot p
        [q \cdot \epsilon  p.\epsilon^{*} - \epsilon \cdot \epsilon^{*} q \cdot p]
      \\ & +2m^{2}_{l}(k_{1} - k_{2} )\cdot p
      \\ & \times [\epsilon \cdot \epsilon^{*}(k_{1} - k_{2} )\cdot p + (k_{1}  - k_{2} ) \cdot \epsilon (p\cdot\epsilon^{*})]
      \\ & +8 [(k_{2} \cdot p)(k_{1}\cdot \epsilon)  - (k_{1} \cdot p)(k_{2} \cdot \epsilon)]
      \\ & \times [(k_{2} \cdot p)(k_{1}\cdot \epsilon^{*})  - (k_{1} \cdot p)(k_{2} \cdot \epsilon^{*})] ,
\end{split}
\end{equation}
where $m_V$ and $m_l$ are masses of vector and pseudoscalar mesons.
The differential decay width of $V \to P l^{+} l^{-}$  is obtained as
\begin{equation}
\begin{split}
d\Gamma(V\to Pl^{+}l^{-})= ~ & \frac{1}{(2\pi)^{5}} \frac{1}{16m_{V}^{2}} |T(\psi(\Upsilon) \to P l^{+} l^{-})|^{2} \\ & \times {\bf |k^{*}|}{\bf |p_{3}|} dm_{l^{+}l^{-}} d\Omega_{3}d\Omega^{*}_{1},
\end{split}
\end{equation}
where $\bf |k^{*}|$ is the momentum of $l^{+}$ or $l^{-}$ in the rest frame of $l^{+} l^{-}$ system, $\bf |p_{3}|$ momentum of the pseudoscalar meson $P$ in the rest frame of  $V$, $d\Omega_{3} = d \phi_{3}d(\cos\theta_{3})$ is the solid angle of $P$ in the rest frame of $V$, and $d\Omega^{*}_{1} = d \phi^{*}_{1}d(\cos\theta^{*}_{1})$ the solid angle of $l^{+}$ or $l^{-}$ in the rest frame of $l^{+} l^{-}$ system (the $z$ direction is defined as the momentum direction of $l^{+} l^{-}$ system in the rest frame of $V$).

\subsection{Differential partial widths for polarized  $\psi$/$\Upsilon$  EM Dalitz decays  }

\begin{figure}[!htbp]
\centering
\includegraphics[width=0.38\textwidth]{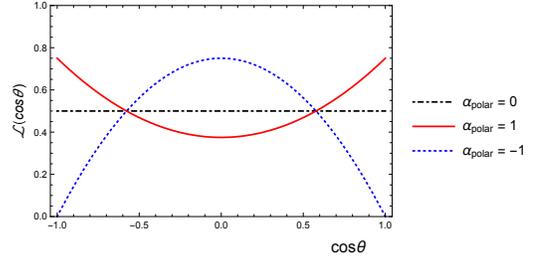}
\caption{
    The angular distributions of $\psi(\Upsilon) \to P l^{+} l^{-}$ with
    different $\alpha_{\rm{polar}}$ values, where the black dot-dashed line
    refers to $\alpha_{\rm{polar}}=0$ (nonpolarization); the red solid curve
    refers to $\alpha_{\rm{polar}}=1$ (transverse polarization); and the blue
    dotted curve refers to $\alpha_{\rm{polar}}=-1$ (longitudinal polarization). 
    }
\label{fig:angle}
\end{figure}

There are totally three  polarization states for the massive vector mesons, which are defined as
\begin{equation}
\begin{split}
& \epsilon^{\mu}_{\rm{LP}} =(0,0,0,1)
\\ & \epsilon^{\mu}_{\rm{TL}} =\frac{1}{\sqrt{2}}(0,1,-i,0)
\\ & \epsilon^{\mu}_{\rm{TR}} =\frac{1}{\sqrt{2}}(0,1,i,0)
\end{split}
\end{equation}
where $\epsilon^{\mu}_{\rm{LP}}$,  $\epsilon^{\mu}_{\rm{TL}}$ and $\epsilon^{\mu}_{\rm{TR}}$  are  the longitudinal polarization,  the left-hand transverse polarization and the right-hand transverse polarization component, respectively. Correspondingly, it is generally known that the angular distribution for each component that has the following form
\begin{equation}
\begin{split}
& \frac{d\Gamma(\psi(\Upsilon) \to Pl^{+}l^{-})_{\rm{LP}}}{d\cos\theta} \sim  1- \cos^{2}\theta
\\ & \frac{d\Gamma(\psi(\Upsilon) \to Pl^{+}l^{-})_{\rm{TL}}}{d\cos\theta} \sim \frac{1+ \cos^{2}\theta}{2}
\\ & \frac{d\Gamma(\psi(\Upsilon) \to Pl^{+}l^{-})_{\rm{TR}}}{d\cos\theta} \sim \frac{1+ \cos^{2}\theta}{2},
\end{split}
\label{equ:angle}
\end{equation}
where $\theta=\theta_3$ is the polar angle of the pseudoscalar $P$ in the rest frame of $\psi(\Upsilon)$.
Therefore, one can determine the polarization of $\psi(\Upsilon)$ by a likelihood fit to the $\cos\theta$ distribution
\begin{equation}
\mathcal{L}(\cos\theta) = 1 + \alpha_{\rm{polar}} \cos^{2}\theta,
\end{equation}
where nonzero $\alpha_{\rm{polar}}$ means partial polarization of
$\psi(\Upsilon)$, $\alpha_{\rm{polar}} = +1$ means only transverse polarization
of $\psi(\Upsilon)$, and $\alpha_{\rm{polar}} = -1$ means only longitudinal
polarization of $\psi(\Upsilon)$, respectively. We show the angular
distributions of $\psi(\Upsilon) \to P l^{+} l^{-}$ with different
$\alpha_{\rm{polar}}$ values in Fig.~\ref{fig:angle}. In $\psi(\Upsilon) \to P
l^{+} l^{-}$, an explicit value of $\alpha_{\rm{polar}}$ from experiment
indicates the polarization information of $\psi(\Upsilon)$.

\subsection{Polarization of  $\psi(\Upsilon)$ in $e^{+} e^{-} \to \psi(\Upsilon)$}
In this section we study the production rate for each polarization state of the vector mesons $\psi$ or $\Upsilon$ in electron-positron colliders. In electron-positron collisions, such as BESIII and Belle II experiments, the amplitude squared of $e^{+} e^{-} \to \psi $ or $\Upsilon$ can be written as
\begin{equation}
\begin{split}
|T(\psi)|^{2} = ~ & \frac{16\pi^{2}\alpha^{2}e^{2}_{c}}{q^{4}}|f_{\psi}|^{2}m^{2}_{\psi}\epsilon^{*}_{\mu}\epsilon_{\nu} \\ & \times (k^{\mu}_{1}k^{\nu}_{2} + k^{\mu}_{2}k^{\nu}_{1} - g^{\mu\nu}k_{1}\cdot k_{2} + g^{\mu\nu}m^{2}_{e}),
\end{split}
\end{equation}
where $e_{c}$ is the electrical charge of the charm quark, $f_{\psi}$ the form factor of $c \bar{c} \to \psi$, $\epsilon_{\mu}$ the polarization four vector of $\psi$, $m_{\psi}$ and $m_e$ the masses of $\psi$ and electron, $q = k_{1}+k_{2}$ with $k_{1}$ and $k_{2}$ the momenta of $e^{+}$ and $e^{-}$, respectively. Then, the relative probabilities for three polarization states can be obtained
\begin{equation}
\begin{split}
|T( \psi)|^{2}_{\rm{LP}}:|T(\psi)|^{2}_{\rm{TL}}:|T( \psi)|^{2}_{\rm{TR}} & = m^{2}_{e}:m^{2}_{\psi}:m^{2}_{\psi} \\ &  \approx 2.7 \times 10^{-8} :1 : 1 .
\end{split}
\end{equation}
The above result shows that the longitudinal polarization can be neglected and
$\psi$ is totally polarized in transverse state in unpolarized $e^+e^-$
collider, which is consistent with the conclusion in Ref. \cite{Richman:1984gh}.
It is easy and intuitive to extend the conclusion to $\Upsilon$ case.

\subsection{Decay rate of $\psi$($\Upsilon$) in $\psi(\Upsilon) \to \eta_c (\eta_b) l^{+} l^{-}$  }
To remove most part of the uncertainty caused by the form factor
$f_{V\!P}(q^2)$, one can consider the ratio of the decay widths of
$\psi(\Upsilon) \to P l^{+} l^{-}$ and $\psi(\Upsilon) \to P \gamma$. The
$q^{2}$-dependent differential decay width of $\psi(\Upsilon) \to P l^{+} l^{-}$
normalized to the width of the corresponding radiative decay $\psi(\Upsilon) \to
P \gamma$ has been given in Ref.~\cite{Fu:2011yy}; here for clearness we quote
it in the following:
\begin{equation}
\frac{d\Gamma(\psi(\Upsilon)\to Pl^{+}l^{-})}{dq^{2}\Gamma(\psi(\Upsilon) \to P \gamma)} = |F_{V\!P}(q^{2})|^{2} \times [\rm{QED}(q^{2})],
\label{equ:calVPll}
\end{equation}
where the normalized transition form factor for the $\psi(\Upsilon) \to P$
transition is defined as $F_{V\!P}(q^{2}) \equiv f_{V\!P}(q^{2})/f_{V\!P}(0)$,
and $\rm{QED}(q^{2})$ represents the QED calculations for point like particles,
\begin{equation}
\begin{split}
\rm{QED(q^{2})} = & \frac{\alpha}{3\pi} \frac{1}{q^{2}} \left(1-\frac{4m^{2}_{l}}{q^{2}} \right )^{\frac{1}{2}} \left(1 + \frac{2m^{2}_{l}}{q^{2}} \right)  \\ & \times \left[ \left(1 + \frac{q^{2}}{m^{2}_{V} - m^{2}_{P}} \right)^{2}  -  \frac{4m^{2}_{V}q^{2}}{(m^{2}_{V} - m^{2}_{P})^{2}} \right]^{\frac{3}{2}}.
\end{split}
\end{equation}
In experiment, by comparing the measured spectrum of the lepton pair in the EM
Dalitz decay with the the QED calculation for pointlike particle, one can
determine the transition form factor in the timelike region of the momentum
transfer~\cite{Landsberg:1986fd}. Namely, the transition form factor can modify
the lepton spectrum as compared with that obtained for pointlike particles.

To estimate the partial width of the $\psi(\Upsilon)$ EM Dalitz decay, the
vector dominance model (VDM) is adopted, in which the hadronic EM current is
proportional to vector meson fields~\cite{GellMann:1961tg, Bauer:1977iq}. Hence,
the transition form factor can be parametrized in the simple pole approximation
\begin{equation}
\label{fractor}
F_{V\!P}(q^{2}) = \frac{1}{1-\frac{q^{2}}{\Lambda^{2}}},
\end{equation}
where the pole mass $\Lambda$ should be the mass of the vector resonance near
the energy scale of the decaying particle according to the VDM model. In $\psi
\to \eta_{c} l^{+} l^{-}$ decays, the pole mass  for $J/\psi$ and
$\psi^{\prime}$ could be the mass of $\psi^{\prime}$ and $\psi(3770)$,
respectively. Similarly, in $\Upsilon \to \eta_{b} l^{+} l^{-}$ decays, the pole
mass for $\Upsilon(\rm{2S})$ ($\Upsilon(\rm{3S})$) decay could be the mass of
$\Upsilon(\rm{3S})$ ($\Upsilon(\rm{4S})$).

\begin{figure}[!htbp]
\centering
\includegraphics[width=0.38\textwidth]{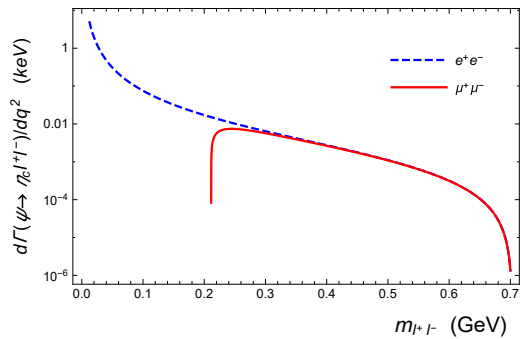}
\caption{The differential decay rates for $\psi^{\prime} \to \eta_c l^{+} l^{-}$, where the red solid curve is  for $\psi^{\prime} \to \mu^+ \mu^-$, the blue dashed curve for $\psi^{\prime} \to \eta_c e^+ e^-$. }
\label{fig:parW}
\end{figure}

By assuming the simple pole approximation, in which we take $\Lambda =
m_{\psi^{\prime}}, m_{\psi(3770)}, m_{\Upsilon(\rm{3S})}, m_{\Upsilon(\rm{4S})}$
for $J/\psi$, $\psi^{\prime}$, $\Upsilon(\rm{2S})$ and $\Upsilon(\rm{3S})$ EM Dalitz
decays, respectively; the partial decay widths of $\psi(\Upsilon) \to
\eta_{c}(\eta_{b}) l^{+} l^{-}$ are estimated and presented in
Table~\ref{tab:VPll}. It is shown that the decay rates for $\psi(\Upsilon) \to
\eta_{c}(\eta_{b}) l^{+} l^{-}$ when $l^{+} l^{-}$ being $\mu^{+} \mu^{-}$ are
approximately one-order smaller then the relevant case when the lepton pair is
$e^{+} e^{-}$, which is due to the suppression of the phase space and the
fast decrease of partial decay rate  as the $q^{2}$ raises shown in Fig. \ref{fig:parW}.
\begin{table}[!htbp]
  \centering
  \caption{The estimated partial decay widths of  $\psi \to \eta_{c} l^{+} l^{-}$  and  $\Upsilon \to \eta_{b} l^{+} l^{-}$ based on Eq.~(\ref{equ:calVPll}) by assuming the simple pole approximation according to VDM model. Here we take $\Lambda = m_{\psi^{\prime}}, m_{\psi(3770)}, m_{\Upsilon(\rm{3S})}$ and $m_{\Upsilon(\rm{4S})}$ for $J/\psi, \psi^{\prime}, \Upsilon(\rm{2S})$ and $\Upsilon(\rm{3S})$ EM Dalitz decays, respectively. The uncertainties are from the errors on measured $\Gamma(\psi \to \eta_{c} \gamma)$ and $\Gamma(\Upsilon \to \eta_{b} \gamma)$.}
  \begin{tabular}{p{2.5cm}p{2.9cm}<{\centering}p{2.9cm}<{\centering}}
  \hline
  \hline
   Decay mode  &  $\Gamma^{VDM}_{ e^{+}e^{-}}$ (keV)  &  $\Gamma^{VDM}_{ \mu^{+}\mu^{-}}$ (keV) \\
  \hline
   $J/\psi  \to \eta_{c} l^{+}  l^{-}$  &  $(9.6 \pm 2.3) \times 10^{-3}$   &  -\\
   $\psi^{\prime}  \to \eta_{c} l^{+}  l^{-}$   &  $(8.9 \pm 1.3)  \times 10^{-3}$  &  $(8.2\pm1.2) \times 10^{-4} $    \\
   $\Upsilon({\rm{2S}}) \to \eta_{b} l^{+} l^{-}$   &  $(10.8 \pm 4.2) \times 10^{-5}$  &  $(8.2 \pm 3.2) \times 10^{-6}$    \\
   $\Upsilon({\rm{3S}}) \to \eta_{b} l^{+} l^{-}$   &  $(9.7\pm1.6) \times 10^{-5}$  &  $(12.6\pm2.1) \times 10^{-6}$    \\
  \hline
  \hline
  \end{tabular}
  \label{tab:VPll}
\end{table}

To study the dependence of the decay rates on the value of the pole mass, we
varied the pole mass. The numerical result shows both the differential and the
total decay rates are not sensitive to the value of the pole mass. The reason
can be well understood. The dominant contribution to the decay rate comes from
the region of the small value of $q^{2}$ for the sake of phase space suppression
in the region of large $q^2$ . For the precesses considered in this work, the
pole masses are larger then the maximum value of $q^2$; therefore, 
$q^{2}/\Lambda^{2}$ is small, thus this term cannot give large effect. Since the
decay rates are not sensitive to the pole mass in the transition form factor,
the estimated partial decay widths in Table~\ref{tab:VPll}  based on the VDM are
reliable.

%\begin{figure}[!htbp]
%\centering
%\includegraphics[width=0.38\textwidth]{figure/Brlambda.eps}
%\caption{The decay rate for $\psi^{\prime} \to \eta_{c} \mu^{+} \mu^{-}$ with the variation of the pole mass $\Lambda$.}
%\label{fig:Lambda2}
%\end{figure}

%===========================================================================
%===========================================================================
%===========================================================================
\section{Discussion}
\subsection{Model predictions of $\psi(\Upsilon) \to \eta_{c}(\eta_{b}) \gamma$ }
\begin{figure}[!htbp]
\centering
\includegraphics[width=0.38\textwidth]{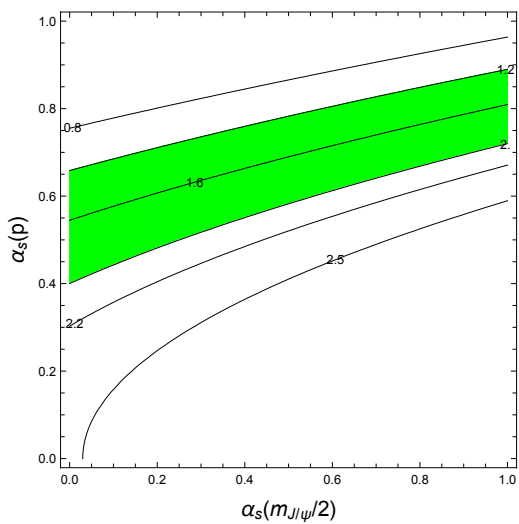}
\caption{
    The contour line of $\Gamma(J/\psi \to \eta_{c}\gamma)$ with
    $\alpha_{s}(m_{J/\psi}/2)$ and $\alpha_{s}(p_{J/\psi})$ from
    Eqs.~(\ref{equ:GNRQCD}) and (\ref{equ:GNRQCDpsi}). The green shadow
    represents the result in experiment, $\Gamma(J/\psi \to
    \eta_{c}\gamma)_{\rm{EXP}}= (1.6\pm0.4)$. 
    }
\label{fig:QCD}
\end{figure}

The EM Dalitz decays, $\psi(\Upsilon) \to \eta_{c}(\eta_{b}) l^{+} l^{-}$, are
related to the radiative decays,  $\psi(\Upsilon) \to \eta_{c}(\eta_{b})
\gamma$, by the transition form factor $f_{V\!P}(q^{2})$ and $f_{V\!P}(0)$.
Models describing $\psi(\Upsilon) \to \eta_{c}(\eta_{b}) \gamma$ can provide
information for $f_{V\!P}(q^{2})$. In Ref.~\cite{Brambilla:2005zw}, using the
theories of NRQCD and pNRQCD, one studied the M1 transitions between two heavy
quarkonia and obtained
\begin{equation}
\Gamma(\psi(\Upsilon) \to \eta_{c}(\eta_{b}) \gamma) = \frac{16e^{2}_{Q}}{3} \frac{\alpha(m^{2}_{\psi(\Upsilon)} - m^{2}_{\eta_{c}(\eta_{b})})^{3}}{8m^{3}_{\psi(\Upsilon)}} w(\alpha_{s}),
\label{equ:GNRQCD}
\end{equation}
where $e_{Q}$ is the electrical charge of the heavy  quark ($e_{c} = 2/3$,
$e_{b}=-1/3$), and $w(\alpha_{s})$ the function of the strong coupling constant
$\alpha_{s}$. The predicted result for $J/\psi \to \eta_{c} \gamma$ is
consistent with data~\cite{Brambilla:2005zw}. However, the predicted results for
$\psi^{\prime} \to \eta_{c} \gamma$ and $\Upsilon(2S) \to \eta_{b} \gamma$ are
larger than data by 2 and 1 order of magnitude, respectively. For
$\Upsilon(3S) \to \eta_{b} \gamma$, the prediction of the branching ratio in the
relativistic quark model is $Br(\Upsilon(3S) \to \eta_{b} \gamma)=4.0\times
10^{-4}$ in Ref.~\cite{Ebert:2002pp}, which is approximately consistent with
experimental measurement $Br(\Upsilon(3S) \to \eta_{b} \gamma)_{\rm EXP}=(5.1\pm
0.7)\times 10^{-4}$ \cite{Tanabashi:2018oca}. The prediction from the
light-front quark model depends on the model-parameters which can be $(1.0\sim
2.5)\times 10^{-4}$ \cite{Ke:2010pp}. This can also be consistent with
experimental data after taking into account both experimental and theoretical
errors.  For $J/\psi \to \eta_{c} \gamma$, the $w(\alpha_{s})$ has the following
form~\cite{Brambilla:2005zw}
\begin{equation}
w(\alpha_{s}) = 1+ C_{F} \frac{\alpha_{s}(m_{J/\psi}/2)}{\pi}  - \frac{2}{3}\left(C_{F}\alpha_{s}(p_{J/\psi}) \right)^{2},
\label{equ:GNRQCDpsi}
\end{equation}
where $C_{F} = 4/3$ is the color coefficient, $\alpha_{s}(m_{J/\psi}/2)$ and $\alpha_{s}(p_{J/\psi})$ represent the values of $\alpha_{s}$  in corresponding energy scale, respectively. Typically, $\alpha_{s}(p_{J/\psi})$ satisfies $p_{J/\psi} \approx m_{c}C_{F}\alpha_{s}(p_{J/\psi})/2 \approx 0.8~\rm{GeV}$~\cite{Brambilla:2005zw}. To study the dependence of $\Gamma(J/\psi \to \eta_{c}\gamma)$ on the value of $\alpha_{s}(m_{J/\psi}/2)$ and $\alpha_{s}(p_{J/\psi})$ , we varied the $\alpha_{s}(m_{J/\psi}/2)$ and $\alpha_{s}(p_{J/\psi})$. The results are shown in Fig.~\ref{fig:QCD}. In this way, a precision measurement of the  partial width of $J/\psi \to \eta_{c}\gamma$ can be used to test QCD model and provide stringent restriction on  $\alpha_{s}$ in charm energy scale.

\subsection{\boldmath  Discussion on the absolute measurement of  the partial widths of $\psi(\Upsilon) \to \eta_{c}(\eta_{b}) l^+l^-$}
%from Eq.~(\ref{equ:GVPG})
With the explicit form factor $f_{V\!P}(q^{2})$, one can obtain the partial
decay widths of $\psi(\Upsilon) \to \eta_{c}(\eta_{b}) \gamma$ where $q^2=0$.
Hence, it is important to measure the $q^2$-dependent form factor
$f_{V\!P}(q^{2})$ , since it can be used to determine the the partial decay
widths of $\psi(\Upsilon) \to \eta_{c}(\eta_{b}) \gamma$ at $q^2 = 0$.

Actually, for the  $\psi(\Upsilon) \to \eta_{c}(\eta_{b}) l^{+} l^{-}$ decay in
$e^+e^-$ collisions at BESIII \cite{Asner:2008nq}, Belle ~\cite{Kou:2018nap}, and
BaBar\cite{Lees:2011mx} experiments, several hundred million $\psi$
($\Upsilon$) are collected.
The lepton usually could be clearly distinguished from pion, kaon, and proton,
so one need only to reconstruct the lepton pair and then look at the recoiling mass
of the lepton pair to obtain the signal. 
In this way, the results are irrelevant to the decay modes of
$\eta_{c}(\eta_{b}) $. Hundreds of million of $\Upsilon(2S)$ and $\Upsilon(3S)$
events are collected by Belle and BABAR detector; we estimate that about
30 signal events could be observed for $\eta_{b} e^{+}e^{-}$ mode with typical
signal efficiency 4\%. 
Since there are more than 10 billion and 800 million $J/\psi$ and $\psi \prime$ 
events collected by BESIII detector up to date, we expect $10^{5}$ and $10^{3}$
signal events from $J/\psi$ and $\psi \prime$, respectively, could be observed
by BESIII with typical signal efficiency 10\%.
So it is possible to measure the branching fraction precisely and probe the
transition form factor.

\section{\boldmath Summary}
In summary, the EM Dalitz decays, $\psi(\Upsilon) \to \eta_{c}(\eta_{b}) l^{+}
l^{-}$, are studied in this work. We investigate the effect of polarization of
$\psi(\Upsilon)$ and estimate the partial decay widths of $\psi(\Upsilon) \to
\eta_{c}(\eta_{b}) l^{+} l^{-}(l=e,\mu)$ by assuming the simple pole
approximation. We obtain the polarization components for the vector mesons
$\psi$ and/or $\Upsilon$ produced in $e^+e^-$ colliders. We find the transverse
polarization states for $\psi$ and/or $\Upsilon$ dominate in unpolarized
$e^+e^-$ colliders. We obtain the angular distribution of $\psi(\Upsilon) \to
\eta_{c}(\eta_{b}) l^{+} l^{-}$ decays for each polarization state of $\psi$
and/or $\Upsilon$, which is helpful to determine the polarization of
$\psi(\Upsilon)$ in the EM Dalitz decays $\psi(\Upsilon) \to \eta_{c}(\eta_{b})
l^{+} l^{-}$ in experiment. The decay widths of  $\psi(\Upsilon) \to
\eta_{c}(\eta_{b}) l^{+} l^{-}$ are also obtained. Besides, we discuss a QCD
model where $\Gamma(J/\psi \to \eta_{c}\gamma)$ is related to $\alpha_{s}$ and
suggest that absolute partial decay widths of EM Dalitz decays should be
measured with reconstruction of only lepton pairs by taking advantage of the
$e^+e^-$ collision with known initial four momentum of the electron and positron
beams.

\section{\boldmath Acknowledgments}
This work is supported in part by the National Natural Science Foundation of
China under Contracts No. 11335009, No. 11875054, and No. 11875168; the Joint
Large-Scale Scientific Facility Funds of the NSFC and CAS under Contract No.
U1532257; CAS under Contract No.QYZDJ-SSW- SLH003; and the National Key Basic
Research Program of China under Contract No. 2015CB856700.

\end{document}